\documentclass[7pt,rmp,aps,preprint,numbers]{revtex4}

\usepackage{graphics,graphicx} 
\usepackage{epsfig}    
\usepackage{amsmath,amssymb}
\usepackage{url}
\usepackage{color}
\usepackage{hyperref}
  \renewcommand{\figurename}{{\bf \footnotesize Figure}\footnotesize}

\usepackage{color}
\usepackage{epsfig}
\definecolor{darkgreen}{rgb}{0,0.5,0} 
\definecolor{violet}{rgb}{0.5,0,0.5}
\definecolor{orange}{rgb}{0.2,0.5,0.5}

\begin{document}

\title{Growth dynamics and the evolution of cooperation in microbial populations}

\author{Jonas Cremer, Anna Melbinger \& Erwin Frey}

\affiliation{Arnold Sommerfeld Center for Theoretical Physics (ASC) and Center for NanoScience (CeNS), Department of Physics, Ludwig-Maximilians-Universit\"at M\"unchen, Theresienstrasse 37, D-80333 M\"unchen, Germany}

\begin{abstract} 
Microbes providing public goods are widespread in nature despite running the risk of being exploited by free-riders. However, the precise ecological factors supporting cooperation are still puzzling. Following recent experiments, we consider the role of population growth and the repetitive fragmentation of populations into new colonies mimicking simple microbial life-cycles. Individual-based modeling reveals that demographic fluctuations, which lead to a large variance in the composition of colonies, promote cooperation. Biased by population dynamics these fluctuations result in two qualitatively distinct regimes of robust cooperation under repetitive fragmentation into groups. First, if the level of cooperation exceeds a threshold, cooperators will take over the whole population. Second, cooperators can also emerge from a single mutant leading to a robust coexistence between cooperators and free-riders. We find frequency and size of population bottlenecks, and growth dynamics to be the major ecological factors determining the regimes and thereby the evolutionary pathway towards cooperation.
 \end{abstract}
\maketitle


One pivotal question in evolutionary biology is the emergence of cooperative traits and their sustainment in the presence of free-riders~\cite{Hamilton64,axelrod-1981-211,Maynard1995,Frank1998,NowakCooperation,West:2010p3514}. By providing a public good, cooperative behavior of every single individual would be optimal for the entire population. However, non-contributing free-riders may take evolutionary advantage by saving the costs for providing the benefit and hence jeopardize the survival of the whole population. In evolutionary theory kin selection \cite{Hamilton64,Michod:1982p1244,Foster:2006p1501,West2007}, multi-level selection \cite{Hamilton:1975,WILSON:1975p2654,Okasha,NowakWilson}, and reciprocity \cite{Trivers} have been found to provide conceptual frameworks to resolve the dilemma~\cite{Frank1998,NowakCooperation,West:2010p3514}. For higher developed organisms, stable cooperation is generally traced back to specific mechanisms like repeated interaction~\cite{Trivers, axelrod-1981-211}, punishment~\cite{Yamagishi, Fehr}, and kin discrimination~\cite{Hamilton64, Queller:2003,West:2006p1249,West:2010p3514}. But how can cooperation emerge in the first place and be maintained without abilities like memory or recognition? Answering this question is especially important within the expanding field of biofilm formation~\cite{ Rainey2003, Velicer:2003p377,Keller,Stewart:2008,Nadell:2009p1177,Xavier:2011}.  There, a successfully cooperating collective of microbes runs the risk to be undermined by non-producing strains saving the metabolically costly supply of biofilm formation~\cite{Velicer:2003p377, West:2006p1249,Nadell:2009p1177,Xavier:2011}.  Sophisticated social behavior cannot be presumed to explain the high level of cooperation observed in nature and experiments~\cite{Rainey2003, Velicer:2003p377, West:2006p1249, Griffin, Buckling:2007p1055,Ackermann:2008p376,Kummerli:2009,Chuang:2009p1054, Gore:2009p1538,Chuang2,Xavier:2011}. Instead, different forms of limited dispersal, such as spatial arrangements, or fragmentation into groups are essential to resolve the dilemma of cooperation among such microbial organisms~\cite{Hamilton64,NowakSpatial, MaynardSmith:1964p1468}. Indeed, in nature microbes typically life in colonies and biofilms. Remarkable, although details strongly differ from species to species, most microbial populations  follow a life cycle of colony initiation, maturation, maintenance and dispersal leading to new initiation, see e.g.~\cite{Costerton:1987je,OToole:2000ve,Hall-Stoodley:2004,Xavier:2011,McDougald:2011bz}.  Well-studied examples include  \emph{Pseudomonas aeruginosa}~\cite{Sauer:2002ux}, \emph{Escherichia coli}~\cite{Beloin:2008ux}, \emph{Bacillus subtilis}\cite{Lemon:2008ux} and \emph{Myxococcus xanthus}~\cite{Velicer:2009}. Even though such a life-cycle is often complexly regulated e.g. by environmental impacts and including collective behavior of colonies, populations bottlenecks alternating with growth phases are essential components of most microbial life-cycles. Employing simplified setups, recent experiments address the role of population bottlenecks and growth by studying  structured microbial populations of cooperators and free-riders~\cite{Griffin, Buckling:2007p1055, Kummerli:2009,Chuang:2009p1054,Chuang2}. In these setups small founder colonies differing in composition were cultivated in separate habitats. For example, Chuang et al.~\cite{Chuang:2009p1054} used 96-well plates as structured environment with a dilution of synthetically designed E.coli strains where the cooperative strain is producer of a public good provoking antibiotic resistance.  A  microbial life-cycle was generated in the lab by regularly mixing all colonies after a certain time and inoculating new cultures. Under these conditions, an increase in the overall level of cooperation was observed even though free-riders have a growth advantage within every colony. However, the precise conditions under which cooperation is favored are subtle~\cite{MaynardSmith:1964p1468,WILSON:1975p2654,SZATHMARY:1987p5987, Motro:1991,Szathmary:1993,Pfeiffer,Killingback, Foster:2006p1501,Traulsen:2006p320, West2007,Wilson:2008p646,Traulsen:2009,NowakWilson,Chuang2}. A possible theoretical explanation for the observed increase in cooperation is the antagonism between two levels of selection,  as  widely discussed in the literature \cite{Okasha}.  Here, these levels, \emph{intra}- and \emph{inter-group} evolution, arise as population dynamics alternates between independent evolution in subpopulations (groups)  and global competition in a merged well-mixed population. Due to the dilemma of cooperation, free-riders are always better off than cooperators within each group (intra-group evolution). In contrast, on the inter-group level, groups with a higher fraction of cooperators are favored over groups with a lower one.

\begin{figure}[]
\begin{center}
\includegraphics[width=0.45\textwidth]{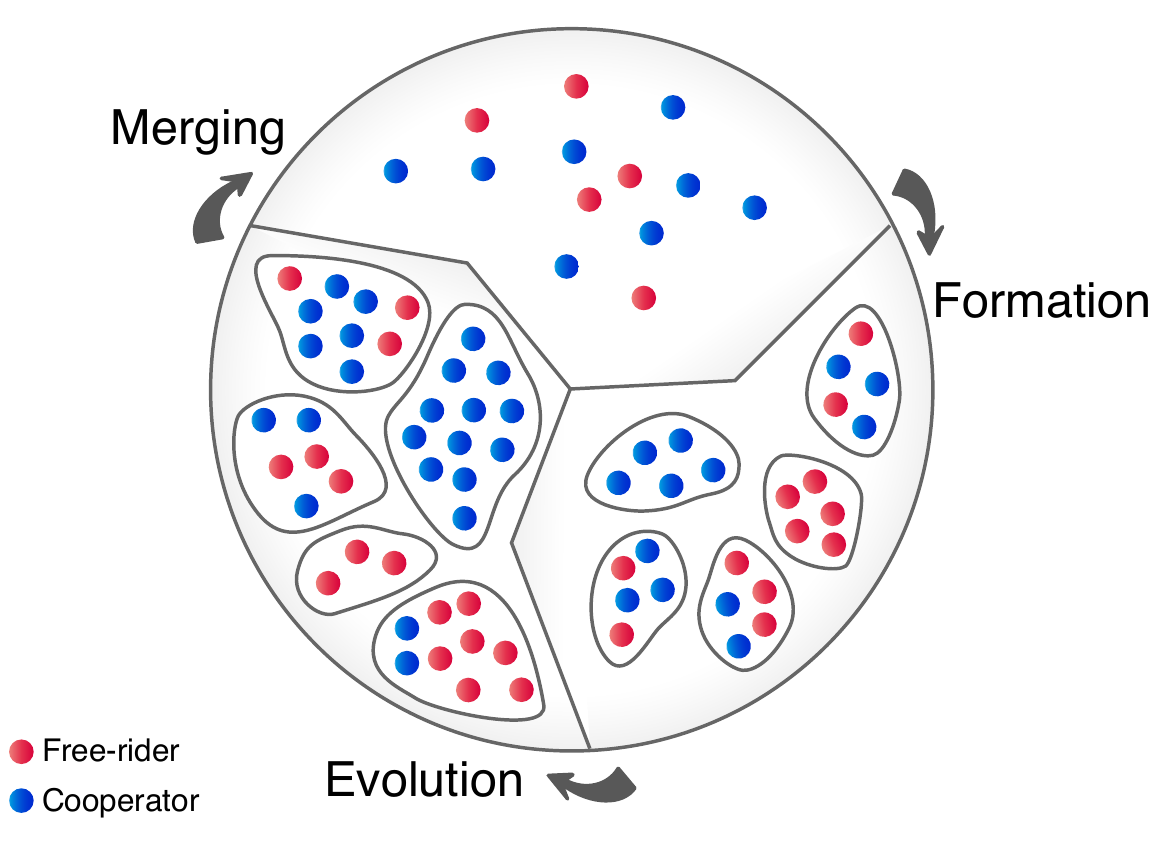}
\end{center}
\caption{{Repetitive cycle of population dynamics.} The time evolution of a population composed of cooperators (blue) and free-riders (red) consists of three cyclically recurring steps. \emph{Group formation step:} we consider a well-mixed population which is divided into $M$ separate groups ($i=1,\ldots,M$) by an unbiased stochastic process such that the initial group size and the fraction of cooperation vary statistically with mean values $n_0$ and $x_0$, respectively.  \emph{Group evolution step:} groups grow and evolve separately and independently; while the fraction of cooperators decrease within each group, cooperative groups grow faster and can reach a higher carrying capacity. \emph{Group merging step:} after a regrouping time, $T$, all groups are merged together again. With the ensuing new composition of the total population, the cycle starts anew. \label{Fig:Cartoon}}
\end{figure}

In this article, we study the interplay between the dynamics at the intra- and inter-group evolution and how it may provoke the maintenance or even the emergence of cooperation. We propose a generic individual-based model which includes three essential elements: a growth disadvantage of cooperators within each group, an advantage of groups incorporating more cooperative individuals,  and regularly occurring regrouping events; cf.~Fig.~\ref{Fig:Cartoon}. Well-known from the  theories of kin \cite{Hamilton64,Michod:1982p1244, West2007,Traulsen:2009} and multi-level selection~\cite{Price,Okasha,Chuang:2009p1054,Traulsen:2009}, cooperation can increase in principle: While, within a group $i$, the fraction of cooperators, $\xi_i$, decreases, groups also change their size, $\nu_i$, such that the fraction of cooperators in the total population, given by the weighted average, $x =\sum_i \nu_i \,  \xi_i \, \large/ \sum_i \nu_i$, may still increase. Such an increase is an example of Simpson's paradox~\cite{Chuang:2009p1054}. To occur, a decreasing fraction of cooperators, $\xi_i$, within groups must be compensated by changing weights, $\nu_i/N$, in the total population of size  $N=\sum_i \nu_i$, i.e.~by a sufficiently high positive correlation between a group's size and its fraction of cooperators~\cite{Price}. Here we want to go beyond stating this mathematical fact and reveal the ecological factors underlying these correlations. To this end the full stochastic dynamics at the intra- and inter-group level will be analyzed. A key element will be the intricate coupling between the dynamics of the composition and the dynamics of the overall size of a group. This applies in particular to microbial populations where the reproduction rate of microbes strongly depends on environmental conditions and thereby also on the composition of the population~\cite{Monod:1949}. Therefore, a proper theoretical formulation has to account for a dynamics in the group size \cite{MelbingerCremer,Cremer:2011a} rather than assuming it to be constant as in most classical approaches \cite{Moran,Fisher,Wright}. Such a dynamic formulation will allow us to investigate ecological mechanisms for the evolution and maintenance  of cooperation.

Motivated by microbial life-cycles~\cite{Costerton:1987je,OToole:2000ve,Hall-Stoodley:2004,Xavier:2011,McDougald:2011bz} and the aforementioned experiments~\cite{Griffin, Buckling:2007p1055,Kummerli:2009, Chuang:2009p1054,Chuang2}, we consider a population of cooperators and free-riders and its evolution in a repetitive cycle consisting of three consecutive steps~\cite{MaynardSmith:1964p1468}, cf.~Fig.~\ref{Fig:Cartoon}. In the \emph{group formation step}, the total population with a fraction of cooperators, $x_0$, is divided into a set of $M$ groups by an unbiased stochastic process such that the group size and the fraction of cooperation vary statistically with mean values $n_0$ and $x_0$, respectively. Subsequently,  the groups evolve independently (\emph{group evolution step}). In each group, both the fraction of cooperators and the group size vary dynamically and change over time. Independent of the specific details, the groups' internal dynamics has the following characteristic features: First, because of the costs for providing the benefit, cooperators have a selection disadvantage, $s$, compared to cheaters in the same group. In particular, cooperators reproduce slower than cheaters and hence the fraction of cooperators decreases within each group (intra-group evolution).  Second, considering the benefit of cooperation, groups with more cooperators grow faster and can reach a higher maximum size (carrying capacity) than groups of mainly cheaters (inter-group evolution)~\cite{MelbingerCremer,Cremer:2011a}; this advantage scales with a parameter, $p$.  These general features can be given a precise mathematical form: Groups follow logistic growth with growth rate and maximum size depending on the fraction of cooperators. For specificity we assume growth conditions comparable to those observed by Chuang et al.~\cite{Chuang:2009p1054}. Details are given in the materials and method section and the supplementary information. After evolving separately for a certain time $t=T$, all groups are merged (\emph{group merging}), and the cycle restarts by forming new groups according to the current fraction of cooperators, $x$, in the whole population. It is the interplay of these three steps, characterized by the initial group size, $n_0$, the selection strength, $s$, and the regrouping time, $T$, which determines the long-term evolution of the population. 

\section*{Results}

\begin{figure}[]
\begin{center}
\includegraphics[width=0.45\textwidth]{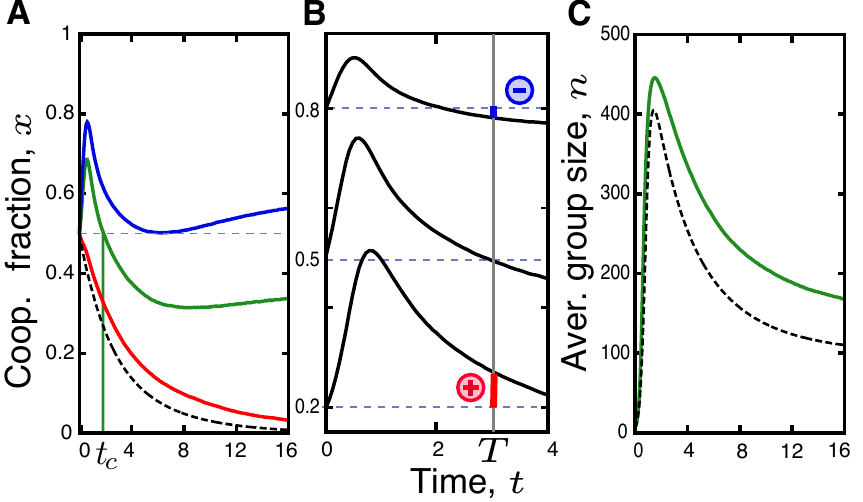}
\end{center}
\caption{{Evolution while individuals are arranged in groups (group-evolution step)}. 
{\bf A} Population average of cooperator fraction, $x$, as a function of time $t$. Depending on the average initial group size, $n_0$, three different scenarios arise: \emph{decrease of cooperation} (red line, $n_0=30$), \emph{transient increase of cooperation} (green line, $n_0=6$, increase until cooperation time $t_c$) and \emph{permanently enhanced cooperation} (blue line, $n_0=4$). These three scenarios arise from the interplay of two mechanisms. While the {\it group-growth mechanism}, due to faster growth of more cooperative groups, can cause a maximum in the fraction of cooperators for short times, the {\it group-fixation mechanism}, due to a larger maximum size of purely cooperative groups, assures cooperation for large times. Both mechanisms become less efficient with increasing initial group sizes and are not effective in the deterministic limit (dashed black line, solution of Eq.~S7). 
{\bf B} The strength of the group-growth mechanism decreases with an increasing initial fraction of cooperators. This is illustrated by comparing the time evolution for three different initial fractions of cooperators and a fixed initial group size $n_0=5$. After a fixed time, here $t=3.03$, the fraction of cooperators is larger than the initial one for $x_0=0.2$, equal to it for $x_0=0.5$, and eventually becomes smaller than the initial value, as shown for $x_0=0.8$. {\bf C} Change of the average group size, $n=\sum_i \nu_i / M$. At the beginning the groups grow exponentially, while they later saturate to their maximum group size. As this maximum size depends on the fraction of cooperators, the average group size declines with the loss in the level of cooperation ($n_0=6$, green line). The deterministic solution (dashed black line) describes this behavior qualitatively. $s=0.1$, $p=10$.
\label{fig:popevol}}
\end{figure}

Fig.~\ref{fig:popevol}{\bf A} shows the time evolution of the overall fraction of cooperators during a group evolution step. We find three distinct scenarios: decrease (red), transient increase (green), and permanent increase of cooperation (blue). Their origin can be ascribed to two ecological mechanisms: more cooperative groups grow faster (\emph{group-growth mechanism}) and purely cooperative groups can reach a larger carrying capacity (\emph{group-fixation mechanism}).

A permanent increase of cooperation can be explained on the basis of the \emph{group-fixation} mechanism: for asymptotically long times the intra-group evolution reaches a stationary state, where each group consist solely of either cooperators or free-riders. Which state is favored depends on the interplay between selection pressure and stochastic effects. Because cheaters have a relative fitness advantage, they tend to outcompete cooperators in groups with a mixed initial composition. However, there are two stochastic effects leading to purely cooperative groups. First, the stochastic process of group formation results in a distribution of group compositions also containing a fraction of groups which consist of cooperators only. Second, random drift \cite{Kimura,Cremer2}, which is most pronounced during a population bottleneck where group sizes are small, can cause a group to become fixed in a state with cooperators only.  Due to the benefit of cooperators for the whole group, these purely cooperative groups reach a much higher carrying capacity than those left without any cooperator. Hence, although inferior in terms of number of groups, purely cooperative groups through their large group size contribute with a large statistical weight to the total composition of the population, and thereby ensure maintenance or even increase of the level of cooperation for long times, cf.~Fig.~\ref{fig:popevol}{\bf A} blue curve.

In order for the group-fixation mechanism to become effective the evolutionary dynamics has to act for time scales longer than the selection time, $t_s := 1/s$, which measures the time scale on which selection acts. For smaller times, a temporary increase in cooperation level is observed provided the initial group size is small enough, cf.~Fig.~\ref{fig:popevol}{\bf A}. The initial rise is caused by the \emph{group-growth mechanism} during the growth phase of colonies, see Fig.~\ref{fig:popevol}{\bf C}. Given a distribution of initial group compositions, it asymmetrically amplifies the size of those groups which contain more cooperators. This effect becomes stronger with a broader distribution, or, equivalently, a smaller initial group size $n_0$. Eventually the initial rise has to decline since, due to the internal selection advantage of free-riders, the fraction of cooperators is always decreasing within each mixed group. As a consequence, the overall benefit of cooperators through faster growth of more cooperative groups is only transient.  After a certain time, the cooperation time, $t_c$, the fraction of cooperators, $x(t)$,  falls  again below its initial value, $x_0$, unless the group-fixation mechanism is strong enough to ensure a permanent increase. Finally, if group-internal selection is too strong compared with the growth advantage of cooperative groups, the level of cooperation cannot increase even transiently, cf.~Fig.~\ref{fig:popevol}{\bf A}, red curve.

\begin{figure}[]
\begin{center}
\includegraphics[width=0.45\textwidth]{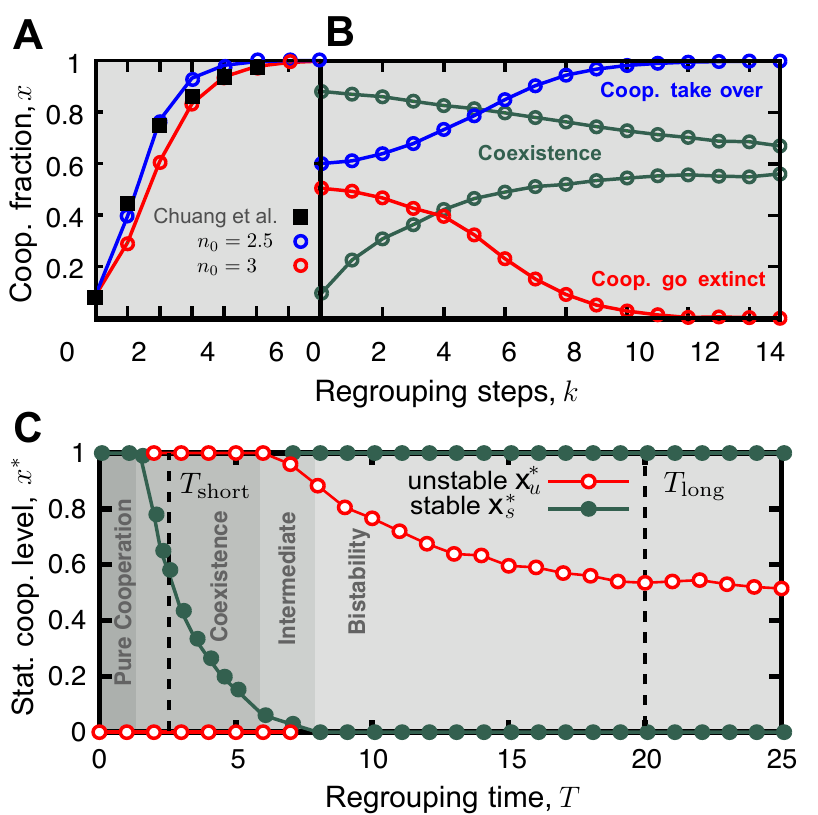}
\end{center}
\caption{Evolution of the overall cooperator fraction under repeated regrouping. After many iterations, $k$, of the evolutionary cycle, a stationary level of cooperation is reached. {\bf A} For small population bottlenecks, $n_0\leq 3$, group-growth and group-fixation mechanisms are effective and lead to purely cooperative populations. Growth parameters, bottleneck size and the regrouping time are chosen according to the experiments by Chuang et al.~\cite{Chuang:2009p1054}, see supplementary information. Without any fitting parameters, our simulation results (colored lines) are in good agreement with the experimental data (black points). {\bf B} For larger bottlenecks, $n_0=5$, and depending on the relative efficiency of the group-growth and group-fixation mechanism, two qualitatively different regimes can be distinguished. While the group-growth mechanism leads to stable coexistence of cooperators and free-riders (green lines), the group-fixation mechanism can lead to a pure state of either only cheaters (red line) or only cooperators (blue line). The relative impact of these mechanisms depends strongly on the regrouping time $T$. For short regrouping times ($T_\text{short} = 2.5 < t_s$, green lines), the group-growth mechanism is effective, while for sufficiently long regrouping times ($T_\text{long} = 20 > t_s$, blue and red lines) the group-fixation mechanism acts more strongly. {\bf C} With parameters equal to {\bf B}, the detailed interplay of the group-growth and group-fixation mechanisms is summarized in a bifurcation diagram showing the stationary levels of cooperation as a function of the regrouping time $T$. Depending of the relative efficiency of both mechanism, four different regimes arise: pure cooperation, coexistence, intermediate, and bistability. The times $T_\text{short}$ and $T_\text{long}$ correspond to the green and red/blue lines shown in {\bf B}. Parameters are  $x_0=0.086$, $T=3.1$, $s=0.05$ and $p=6.6$ in {\bf A}; see also supplementary information. In {\bf B}, $x_0 = \lbrace 0.1 \;(\text{green})$, $x_0 = 0.9\;(\text{green})\rbrace$ and $x_0 = \lbrace 0.5 \;(\text{red}), x_0 = 0.6 \;(\text{blue})\rbrace$ for $T_\text{short} = 2.5$ and $T_\text{long} = 20$, respectively. $s=0.1$ and $p=10$ in {\bf B/C}. 
\label{fig:a-plot}}
\end{figure}

Combining all three steps of the cycle we now ask for the evolutionary outcome after many iterations, $k$, of the cycle. For very small bottlenecks, $n_0\leq 3$, both the group-fixation and the group-growth mechanism result in a a purely cooperative population and cannot be distinguished. This is shown in Fig.~\ref{fig:a-plot}{\bf A} for parameters corresponding to the experiments by Chuang et al.~\cite{Chuang:2009p1054}; the experimental results and the results of our stochastic model are in excellent agreement. For larger bottlenecks, $n_0=5$, and depending on the relative magnitude of the \emph{regrouping time} $T$, we find two fundamentally distinct scenarios, see Fig.~\ref{fig:a-plot}{\bf B}. For large regrouping times, $T \gg t_s$, there is a threshold value, $x_u^*$, for the initial cooperator fraction, $x_0$, above which cooperators take over the whole population and below which they go extinct. In contrast, for regrouping times smaller than the selection time, $T \leq t_s$, independent of the initial value, $x_0$, the population reaches a stationary state where cooperators are in stable coexistence with free-riders. As explained next, these two scenarios are closely tied to the group-growth and group-fixation mechanisms; for an illustration see the supplementary videos.
The threshold value for maintenance of cooperation at large regrouping times is a consequence of  group-fixation and the larger carrying capacity of purely cooperative groups. Since for $T \gg t_s$ the intra-group dynamics has reached a stationary state, fixation leaves the population with groups consisting of either cooperators or defectors only. The probability of fixation in the respective state and hence the fraction of purely cooperative groups after completing one cycle strongly depends on the initial cooperator fraction.  Now, if the initial cooperator fraction becomes too low, the number of cooperative groups will be too rare such that even their larger maximum group size is no longer sufficient for them to gain significant weight in the total population, and the overall cooperator fraction in the population will decline.  Thus there must be a critical value for the cooperator fraction, $x^*_u$, below which, upon iterating the cycle the fraction of cooperators will decline more and more, see  Fig.~\ref{fig:a-plot}{\bf B} (red line). In contrast, above the critical value purely cooperating groups are becoming more frequent upon regrouping, and therefore cooperators will eventually take over the population completely, cf. Fig.~\ref{fig:a-plot}{\bf B} (blue line).

When groups are merged during the phase of transient increase of cooperation,  $T \leq t_s$,  the stationary level of cooperation does not depend on the initial one. This behavior is due to the dependence of the change of the cooperator fraction during one cycle, $\Delta x$, on the initial fraction, $x_0$ as discussed in the following; see also Fig.~\ref{fig:popevol}{\bf B}.  As we have already eluded to in the discussion of the group-growth mechanism, stochasticity during group formation and during the initial neutral phase of the group evolution dynamics results in a broad distribution of group compositions. The evolutionary dynamics is acting on this distribution in an antagonistic fashion. While, due to the higher growth rate of more cooperative groups, the distribution develops a positive skew leading to an increase in the average overall cooperation, the group-internal selection pressure is counteracting this effect by reducing the cooperator fraction within each group. The relative strength of the former effect is largest for small initial cooperator fraction since this allows the largest positive skew to develop. Hence, for a given regrouping time, if the change in overall cooperator fraction $\Delta x$ is positive for small $x_0$ it must become negative for sufficiently large $x_0$, as illustrated in  Fig.~\ref{fig:popevol}{\bf B}. For a more detailed mathematical discussion of these effects we refer to the supplementary information. As a consequence, in populations with a small initial fraction of defectors, the defectors increase in frequency. At the same time, when the initial fraction of cooperators is low, they also increase in number, finally leading to stable coexistence of cooperators and defectors at some fraction $x^*_s$. This stationary fraction of cooperators is independent of the starting fraction and solely determined by the parameters of the evolutionary dynamics. 

The interplay of both the group-growth and group-fixation mechanism leads, depending on the regrouping time, to different scenarios for the levels of cooperation. These are summarized in the bifurcation diagram Fig.~\ref{fig:a-plot}{\bf C}, where the stable and unstable fixed points of the regrouping dynamics, $x_s^*$ and $x_u^*$, are shown as functions of the regrouping time. The scenarios can be classified according to their stability behavior under regrouping as follows: For large regrouping times, $T\gg t_s$, the group-fixation mechanism leads to \emph{bistable} behavior. With decreasing $T$, the fixation mechanism loses ground while the group-growth mechanism becomes more prominent. There is a \emph{intermediate} scenario: the dynamics is bistable with full cooperation as well as coexistence as stable fixed points. For even smaller times, only the group-growth mechanism remains effective and the rare strategy here always outperforms the common one such that each strategy can invade but not overtake the other: \emph{coexistence}. Finally, for $T\ll t_s$, cooperators always take over the population, effectively leading to  \emph{purely cooperative populations}.

\section*{Discussion}
In this article, we have studied the influence of population dynamics and fluctuations on the evolution and maintenance of cooperation. We specifically account for alternating population bottlenecks and phases of microbial growth. Thereby, our model serves as a null-model for cooperation in rearranging  populations~\cite{Griffin, Buckling:2007p1055,Kummerli:2009, Chuang:2009p1054,Chuang2}, e.g. during microbial and parasitic life-cycles~\cite{Strassmann:2000,Velicer:2009,Wilson:1977,Dronamraju,Xavier:2011}, and bacterial biofilm formation~\cite{Costerton:1987je,OToole:2000ve,Hall-Stoodley:2004,Xavier:2011,McDougald:2011bz}. The final outcome of the dynamics depends on the interplay between the time evolution of size and composition of each subpopulation. While a growth advantage of more cooperative groups favors cooperators, it is counteracted by the evolutionary advantage of free-riders within each subpopulation. We have investigated the stochastic population dynamics and the ensuing correlations between these two opposing factors. Depending on whether groups are merged while they are still exponentially growing or already in the stationary phase, two qualitatively different mechanisms are favored, the group-growth and the group-fixation mechanism. Importantly, our analysis identifies demographic noise as one of the main determinants for both mechanisms. First, demographic noise during population bottlenecks creates a broad distribution in the relative abundance of cooperators and free-riders within the set of subpopulations. The growth advantage of more cooperative subpopulations implies an asymmetric amplification of fluctuations and possibly yields to an increase of cooperation in the whole population (group-growth mechanism).  Our analysis shows that this can enable a single cooperative mutant to spread in the population which then, mediated by the dynamics, reaches a stationary state with coexisting cooperators and free-riders. Second, if the founder populations contain only very few individuals, demographic fluctuations strongly enhance the fixation probability of each subpopulation which then consists of cooperators or free-riders only. Purely cooperative groups can reach a much higher carrying capacity. However, only if the relative weight of purely cooperative groups is large enough, this effect leads to an increase in the level of cooperation in the whole population (group-fixation mechanism). From our theoretical analysis of the population dynamics we conclude this to be the case only if the initial fraction of cooperators is above some threshold value.

As shown by comparison with experiments by Chuang et al.~\cite{Chuang:2009p1054} the proposed model is able to describe microbial dynamics quantitatively. Moreover, our model makes predictions how the evolutionary outcome varies depending on population dynamics and bottlenecks, and how the discussed mechanisms can provoke cooperation. These predictions can be tested experimentally by new experiments similar to those of Chuang et al. and others~\cite{Griffin, Buckling:2007p1055,Kummerli:2009, Chuang:2009p1054,Chuang2}: For example, by varying easily accessible parameters like the bottleneck size $n_0$ or the regrouping time $T$, the relative influence of both mechanisms can be tuned. Then the resulting level of cooperation and the ensuing bifurcation diagrams can be quantitatively compared with our theoretical predictions.

As we assume the worst case scenario for cooperators, e.g randomly formed groups and no additional assortment, our findings are robust: The discussed pathways towards cooperation based on a growth-advantage of more cooperative groups and restructuring are expected to stay effective when accounting also for other biological factors like positive assortment, spatial arrangements of groups, mutation, or migration\footnote{publication in preparation}. 

Shown by our analysis, a regular life-cycle favors cooperation. Besides better nutrient exploitation, this advantage for cooperation might be one reason for the evolution of more complex, controlled life-cycles including collective motion of microbes, local lysis, and sporulation~\cite{Costerton:1987je,OToole:2000ve,Hall-Stoodley:2004,Xavier:2011,McDougald:2011bz}.

\section*{Methods}
We used a stochastic, individual-based model where each individual is either a cooperator or a free-rider. In the group formation step groups are formed at random. The initial group size, $\nu_{0,i}$, is Poisson distributed (with mean $n_0$). Given this size, the fraction of cooperators $\xi_{0,i}$ follows by a binomial distributed number of cooperators. During the evolution step, each individual is subject to random birth and death events. The dynamics is given by a time-continuos Markov process where the change of the probability, $\partial_t P(\nu_i,\xi_i;t)$, is given by a master equation. In detail, the basal per capita birth rate of each individual depends linearly on the group level of cooperation $\xi_i$, while the per capita death rate increases linearly with the group size $\nu_i$ the individual belong to. In addition, free-riding individuals have a higher birth-rate where the strength of selection $s$ measures the advantage of free-riding individuals. Full details are given in the supplementary information. The time scale is such that a small population of only free-riders initially grows exponentially with the average size $\nu_{i,0}\exp t$. To investigate the dynamics and both evolutionary mechanisms we performed extensive computer simulations by employing the Gillespie algorithm~\cite{Gillespie}. Group size is $M=5\cdot 10^3$ in Fig.~2, and $M=5\cdot 10^4$ in Fig.~3. 


\vspace{1cm}
\noindent {\bf Acknowledgments}\\
We thank Jan-Timm Kuhr and Matthias Lechner for helpful discussions. Financial support by the Deutsche Forschungsgemeinschaft through the SFB TR12
``Symmetries and Universalities in Mesoscopic Systems" is gratefully acknowledged.\\
{\bf Author contributions}\\
J.C., A.M., and E.F. designed the research. J.C., and A.M. performed the stochastic simulations and analyzed the data. J.C., A.M., and E.F. wrote the manuscript.\\
{\bf Additional information}\\
Competing Financial Interests
The authors declare no competing financial interests.


\renewcommand{\theequation}{S\arabic{equation}}
\renewcommand{\thefigure}{S\arabic{figure}}
\renewcommand{\thetable}{S\arabic{table}}
\renewcommand{\figurename}{Figure}



\newpage

\noindent {\Large Supplementary Information\\ Growth dynamics and the evolution of cooperation in group-structured populations} 
\vspace{1cm}

\noindent{\large Jonas Cremer,$^{1}$ Anna Melbinger,$^{1}$ Erwin Frey$^{1\ast}$

\noindent \normalsize{$^{1}$ Arnold Sommerfeld Center for Theoretical Physics (ASC) and Center for NanoScience (CENS),}
\normalsize{Department of Physics, Ludwig-Maximilians-Universit\"at M\"unchen,}
\normalsize{ Theresienstrasse 37, D-80333 M\"unchen, Germany}\\
\normalsize{$^\ast$To whom correspondence should be addressed; E-mail:  frey@lmu.de.}
}

\vspace{1cm}

\noindent In this supplementary text, we give a more detailed discussion of our model and the group-growth mechanism. Furthermore we show comparisons of our analysis with experiments by Chuang et al.~\cite{Chuang:2009p1054SI}.

\section{The Model}

Here, we give details on the consecutive steps of the "life-cycle" of the meta-population. We first specify the group formation step before considering the dynamics within groups (group evolution step).

\subsection{The Group Formation Step}

Starting with an initial fraction of cooperators $x_0$ in the population, $M$ groups are formed. Both, group size and group composition, are randomly distributed. Each group $i\in [1,M]$ initially consists of $\nu_{0,\,i}$ individuals. If the population from which the groups are formed is much larger than $Mn_0$,  $\nu_{0,\,i}$ follows a Poisson distribution\footnote{This holds for typical conditions of small population bottlenecks as observed in bacterial life-cycles.},
\begin{equation}
P(\nu_{0,\,i})=\frac{n_0^{\nu_{0,\,i}}}{\nu_{0,\,i}!}\exp\left(-n_0\right),
\end{equation}
with mean $n_0$.
Further, the initial composition of each group is also formed randomly. The probability for $\zeta_{0,\, i}$  cooperators in a group $i$ is assumed to be given by a Binomial distribution 
\begin{equation}
P(\zeta_{0,\,i})=\begin{pmatrix}\nu_{0,\,i} \\ \zeta_{0,\,i}\end{pmatrix}x_0^{\zeta_{0,\,i}}(1-x_0)^{\nu_{0,\,i}-\zeta_{0,\,i}}
\end{equation}
with mean $x_0\nu_{0,i}$. The initial fraction of cooperators $\xi_{0, \,i}$ within each group is thereby given by $\xi_{0, \,i}=\frac{\zeta_{0,\,i}}{\nu_{0,\,i}}$.

By this we assume the groups to be formed at random without any bias. This corresponds to a worst case scenario for cooperators which gain no additional advantage due to positive assortment. Note, that the same initial distribution of group compositions is reached if one assumes both, the initial number of cooperators (C) and free-riders (F), to be Poisson distributed with mean values $\lambda_C$ and $\lambda_F$, respectively. The mean values are related by $n_{0}=\lambda_C + \lambda_F$ and $x_0=\lambda_C/(\lambda_C+\lambda_F)$.

\subsection{The Group Evolution Step}
After the groups were formed randomly, they grow and evolve separately. In the following, we consider the dynamics within one specific group $i$ in detail. As emphasized in the main text, we include two essential requirements experiments on microbial systems have in common. First, in each group cooperators ($C$) grow slower than free-riders ($F$). Second, groups with a higher fraction of cooperators grow faster and are bounded by a higher maximum group size (carrying capacity) than groups with a lower one. To account for these facts, the growth rates have to consist of a group related and a trait/type specific part~\cite{MelbingerCremerSI,Cremer:2011aSI}. We,
therefore, denote the per capita growth rate of an individual of type  $S\in\{C,F\}$ within group $i$ as
\begin{equation}
G_S(\xi_i)=g(\xi_i)\frac{f_S(\xi_i)}{\langle f\rangle},\label{eq:birth}
\end{equation}
where $g(\xi_i)$ is the group related, $f_S(\xi_i), ~S\in\{C,F\}$ is the species related part, and $\langle f \rangle=\xi_i f_C(\xi_i)+(1-\xi_i)f_F(\xi_i)$ is the average fitness. The normalization of the fitness, $f_S(\xi_i)/\langle f \rangle$, is a convenient choice to disentangle the influence of global and relative parts more easily.
Further, the group related part, $g(\xi_i)$,  which accounts for the growth advantage of more cooperative groups, is assumed to increase linearly with $\xi_i$. 
For specificity, we use experimental conditions similar to those presented in reference~\cite{Chuang:2009p1054SI,Chuang2SI}. In these experiments, a purely cooperating population growth to an about ten times higher population size than a purely defecting one. In our model, the maximum population size scales with $g$ and therefore we  set 
\begin{eqnarray}
g(\xi_i)= r(1 + p\xi_i).
\end{eqnarray} 
Here $r$ determines the overall time scale for growth and defines our units of time, i.e.\ it is set to one unless specified otherwise. In the main text we have used $p=10$ for specificity; see also section \ref{comparisonchuang} where we compare with the experimental data by Chuang et a. \cite{Chuang:2009p1054SI}.

Note, however, that the qualitative findings, especially both evolutionary mechanisms, do not depend on the exact form of $g(\xi_i)$ but only on the fact that $g(\xi_i)$ is monotonically increasing with the fraction of cooperators. The trait specific part, $f_S(\xi_i)$, includes the different growth rates of cooperators and free-riders within group $i$. We here employ the standard formulation of evolutionary game theory and assume it to be given by the payoff matrix of a Prisoner's dilemma game~\cite{NowakSI,HofbauerSI}. The trait specific parts are given by
\begin{align}
f_C(\xi_i)=&1+s\left[b\xi_i-c\right], \nonumber\\
f_F(\xi_i)=&1+sb\xi_i,
\end{align}
and the fitness advantage of  free-riders $\Delta f=f_F(x)-f_C(x)=-sc$ is frequency independent. For specificity, we set $b=3$ and $c=1$. Thereby, the selection strength $s$ is the only free parameter controlling the fitness difference, $\Delta f$, which corresponds to the advantage of free-riders within each group. In the experiments~\cite{Chuang:2009p1054SI,Chuang2SI}, the selection strength was of the order $s\sim 0.05$. In our manuscript, we set $s=0.1$ as an upper approximation of this value.

To model growth  bounded by restricted resources we further introduce per capita death rates which increase linearly with the number of individuals in a group,
\begin{align}
D_S(\nu_i)=\frac{\nu_i}{K}.\label{eq:death}
\end{align}
These are independent of the specific type $S$ and lead to logistic-like growth within each group. $K$ sets the scale of the maximum group size~\cite{VerhulstSI}. In detail, for purely defecting groups the carrying capacity is $K$ while it is $(1+10)K$ for purely cooperating ones. For the discussed results, only the ratio of group sizes and not their absolute values are important. Hence, for numerical convenience, we set $K$ to a  constant value, $K=100$.

The full stochastic dynamics follows a master equation which can be derived by the per capita growth and death rates, Eqs.~(\ref{eq:birth}) and (\ref{eq:death}). This master equation gives the temporal evolution of $P(\xi_{\,i};\nu_{i};t)$, the probability for group $i$ to consist of $\nu_{i}$ individuals with a fraction of $\xi_i$ cooperators at time $t$. We use the Gillespie algorithm to perform stochastic simulations~~\cite{GillespieSI}.

\begin{figure}
\centering
\includegraphics[width=0.9\textwidth]{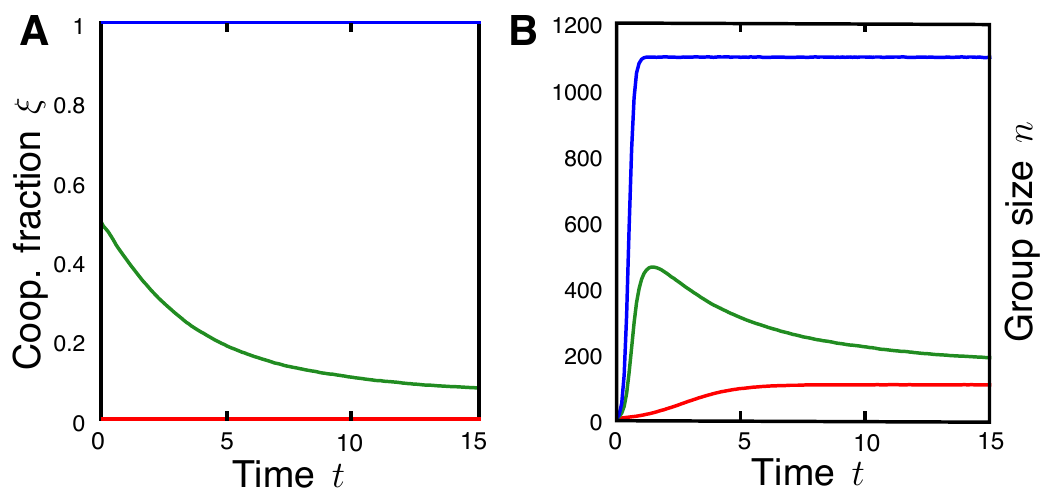}
\caption{Dynamics in single groups. {\bf A} Evolution of cooperation. For a mixed group (green), the fraction of cooperators declines due to the fitness advantage of free-riders while it stays constant for purely cooperating (blue) or defecting (red) groups. {\bf B} Logistic like growth of the group size. For pure groups, the group-related advantage of more cooperative groups is most visible. Purely cooperating groups (blue) grow faster and reach a larger maximum carrying capacity than groups of only free-riders (red). A mixed group (green) grows faster than a group of only free-riders at the beginning. However also in the initially mixed group only free-riders can remain in the long run, and the carrying capacities of both groups become the same. The simulation average over different realizations of only one group. Parameters are $n_0=6$ and $s=0.1$, $\xi_{0}$ is equal to $0$ (red), $0.5$ (green), and $1$ (blue).}\label{fig:S1}

\end{figure}

While fluctuations strongly affect the dynamics,  it is still instructive to look at the deterministic description where fluctuations during the group-evolution step are neglected. This deterministic dynamics within each group, $i$, is then given by rate equations for the fraction of cooperators $\xi_i$ and the total group size $\nu_i$:
\begin{eqnarray}
\partial_t \xi_i&=&-s (1+10 \xi_i)\xi_i(1-\xi_i),\nonumber\\
\partial_t \nu_i&=&(1+10 \xi_i-\nu_i/K)\nu_i.\label{eq:mf}
\end{eqnarray}
Thus, in a deterministic manner, intra-group evolution is described by a replicator-like dynamics while the size of each group follows logistic growth (with a $\xi_i$ dependent growth rate and carrying capacity). We illustrate this dynamics in Fig.~S\ref{fig:S1} for three different initial conditions.

\section{The Group-Growth Mechanism}
As stated in the main text, the group-growth mechanism relies on the faster growth of more cooperative groups. Even though  cooperators reproduce slower compared to free-riders in the same group, the positive effect on group-growth can outbalance this disadvantage. 
For an illustration, see the specific example given in Table~\ref{tab:1}.

\begin{table}
\caption{Per capita growth rates of cooperators and free-riders in two groups}
\begin{center}
\begin{tabular}{|l||c|c|}
\hline
&group 1& group 2\\
\hline\hline
fraction of cooperators $\xi_{i}$  & $3/4$ & $1/4$ \\
\hline
per capita growth rate cooperators, $g(\xi_i)f_C(\xi_i)/\langle f \rangle$ &$8.31$& $3.33$ \\
\hline per capita growth rate free-riders, $g(\xi_i)f_F(\xi_i)/\langle f \rangle $ & $9.05$ & $3.58$ \\
\hline
\end{tabular}
\end{center}
Two groups, $i=1$ and $i=2$ in comparison. While the per capita growth rates of cooperators are smaller than the per capita growth rates of free-riders within every group, the per capita growth rate of cooperators in the more cooperative group $1$ strongly exceeds the per capita growth rate of free-riders in the less cooperative group $2$ due to the group related fitness $g(\xi_i)$. The strength of selection is given by $s=0.1$.\label{tab:1}
\end{table}

Here, we quantify the growth advantage of more cooperative groups. For this, we consider only short times $t\ll 1/s$. Then, and in the limit of weak selection, $s\ll 1$, the deterministic time evolution, given by Eqs.~(\ref{eq:mf}), is
\begin{align}
\xi_i=&\xi_{0,i}\nonumber\\
\nu_i=&\nu_{0,i}\exp\left[g(\xi_{0,i})t\right]\nonumber.
\end{align}
The overall fraction of cooperators can be calculated by averaging over all possible initial group compositions,
\begin{align}
 x(t)=\frac{\sum_{i} P(\xi_{0,\,i};\nu_{0,i})\xi_i\nu_i}{\sum_{i} P(\xi_{0,\,i};\nu_{0,i})\nu_i}\nonumber.
\end{align}
By differentiating with respect to time $t$, we find the following expression
\begin{align}
\frac{d}{dt} x=\text{Cov}(x,g(x))\label{eq:cov}.
\end{align}

This corresponds to a Price equation on the group level \cite{PriceSI,OkashaSI}, here stating that an increase in the fraction of cooperators is possible in principle if there is a positive correlation between $x$ and the group related growth $g(x)$. 
However, for longer times $t>1/s$ the selection advantage of free-riders counteracts the group-growth mechanism such that it can only act transiently.

\begin{figure}
\centering
\includegraphics[width=\textwidth]{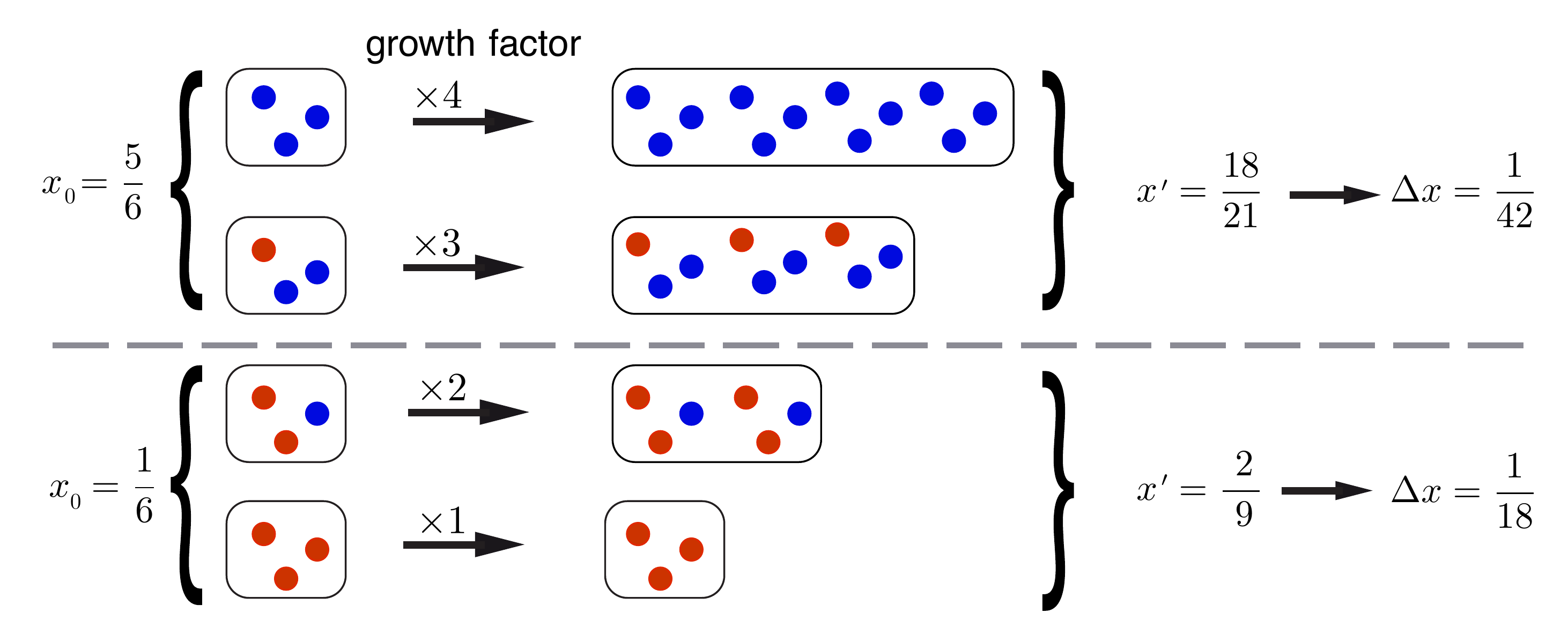}
\caption{How the group-growth mechanism depends on the fraction of cooperators. Two sets of two groups are compared, one with a low fraction of cooperators (bottom) and one with a high one (top). Both groups evolve for a certain time, here with $g\propto1+3x$ and no selection advantage for free-riders, $s=0$. As can be readily seen, the change in the fraction of cooperators is larger for groups with a smaller initial fraction of cooperators.\label{fig:cartoon}}
\end{figure}

As shown in the main text, the strength of the group-growth mechanism depends strongly on the initial fraction of cooperators. This is illustrated in Fig.~S\ref{fig:cartoon}. 

\section{Comparison with experiments on synthetic microbial system by Chuang, Rivoire and Leibler}
\label{comparisonchuang}

\begin{figure}[htb]
\centering
\includegraphics[width=0.9\textwidth]{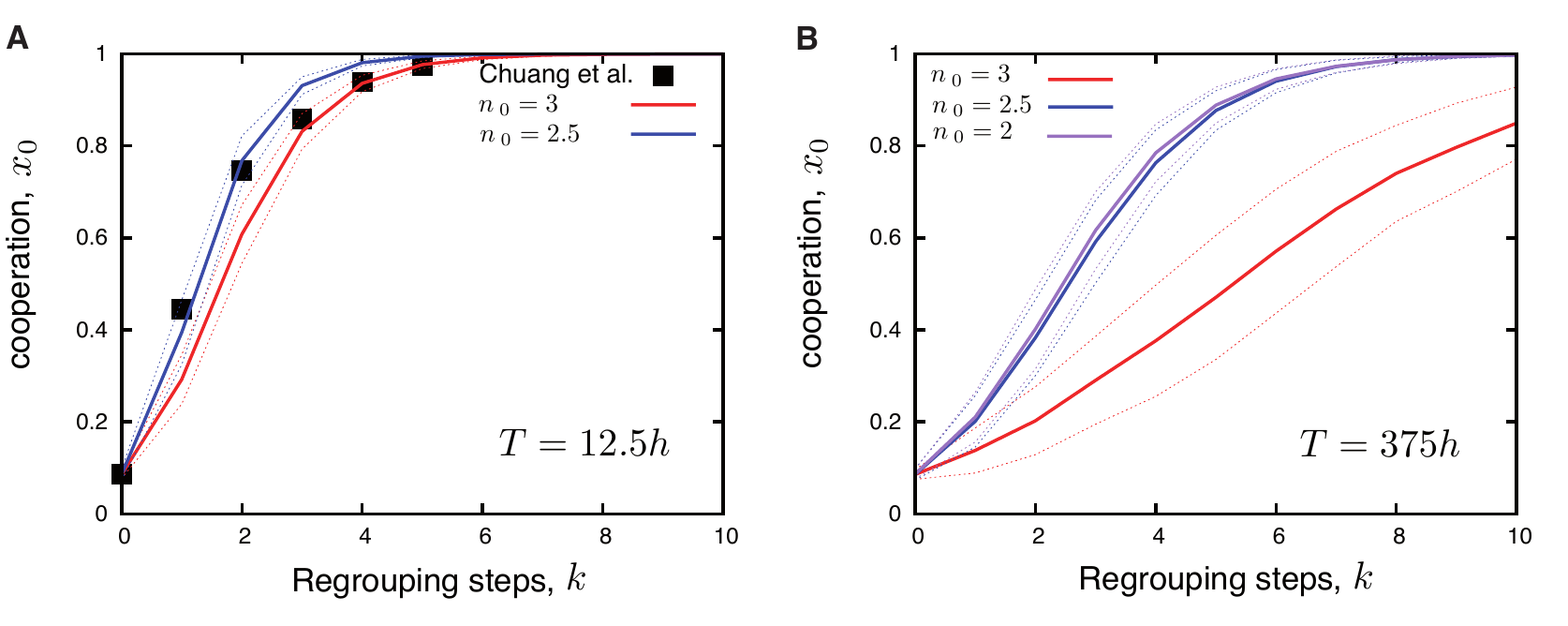}
\caption{Increase in the level of cooperation for conditions resembling those examined by Chuang et al.~\cite{Chuang:2009p1054SI}.
{\bf A} Short regrouping time, $T=12.5h$. The measurements by Chuang et al. (black points) in comparison with the predictions of our model. Solid lines denote the expected level of cooperation. The dashed lines show the corresponding mean plus/minus the standard deviation. {\bf B}, Large regrouping time $T=375h$. For similar conditions, but a longer regrouping time, the outcome is qualitatively the same and only cooperators prevail. For both parts of the figure parameters are  $x_0=0.086$, $r=6.8\times 10^{-4}\, \text{min}^{-1}$, $f_C=1$, $f_D=1.05$, $p=6.6$. In {\bf A},  $K=1.5\times 10^6$. In {\bf B}, $K=1.5\times 10^5$.
\label{fig:comparison_chuang}}
\end{figure}

We have compared our theoretical analysis with recent experiments by Chuang et al. \cite{Chuang:2009p1054SI} on a synthetic microbial model system. They have studied regrouping populations with initial population size $n_0$ in the range between $2$ and $3$, an initial cooperator fraction of $x_0 = 0.086$, and a regrouping time $T = 12-13$~h. Other model parameters were estimated as follows. The inherent fitness advantage of free-riders relative to cooperators was observed to be in the range between $1.04$ and $1.05$. In our model this translates to
\begin{eqnarray}
f_C &=& 1 \, ,\\
f_D &=& 1.05 \, ,
\end{eqnarray}
where in contrast to equation~\eqref{eq:birth} we did not normalize the species related part, i.e.~$\langle f \rangle\equiv 1$.
The growth curves for different compositions of the population (see Fig.S3 in \cite{Chuang:2009p1054SI}) give access to the overall growth rate and its frequency dependence. From Fig.S3 in \cite{Chuang:2009p1054SI} we estimate:
\begin{eqnarray}
r&=& 6.8 \times 10^{-4} \, \text{min}^{-1} \, ,\\
p &=& 6.6 \, .
\end{eqnarray}
Employing these parameters in our model we have simulated the regrouping dynamics and find good agreement with the experimental results, cf. Fig.~\ref{fig:comparison_chuang}a. Since the population dynamics is still within the exponential growth phase at the regrouping time, we interpret the observed increase of cooperation as a group-growth mechanism. However, because of the particular set of experimental parameters, the resulting stationary cooperator fraction is very close to one which makes it difficult to observe coexistence between cooperators and free-riders. We can now use our theoretical model to explore the effects of an increase in the regrouping time. Changing the regrouping time from $T = 12.5$~h to $T = 375$~h we find that the time evolution of the cooperator fraction remains qualitatively similar, despite the fact that now cooperation increases because of the group-fixation mechanism, cf. Fig.~\ref{fig:comparison_chuang}b. Thus even by changing the regrouping time these small values of $n_0$ do not allow to distinguish between the two mechanisms. However, as discussed in the main text, larger values of $n_0$ (in the range of $4-6$) give a clear signature of each of the mechanisms upon varying the regrouping time.

\end{document}